\documentclass[preprint]{aastex}
\shorttitle{Two-sided-loop jets}
\slugcomment{Submitted to ApJ}
\shortauthors{Zheng et al.}
\begin{document}

\title{Two-sided-loop jets associated with magnetic reconnection between emerging loops and twisted filament threads}
\author{Ruisheng Zheng$^{1}$, Yao Chen$^{1}$, Zhenghua Huang$^{1}$, Bing Wang$^{1}$, and Hongqiang Song$^{1}$}
\affil{$^{1}$Shandong Provincial Key Laboratory of Optical Astronomy and Solar-Terrestrial Environment, and Institute of Space Sciences, Shandong University, 264209 Weihai, China; ruishengzheng@sdu.edu.cn\\}

\begin{abstract}
Coronal jets are always produced by magnetic reconnection between emerging flux and pre-existing overlying magnetic fields. When the overlying field is vertical/obilique or horizontal, the coronal jet will appear as anemone type or two-sided-loop type. Most of observational jets are of the anemone type, and only a few of two-sided-loop jets have been reported. Using the high-quality data from New Vacuum Solar Telescope, Interface Region Imaging Spectrograph, and Solar Dynamics Observatory, we present an example of two-sided-loop jets simultaneously observed in the chromosphere, transition region, and corona. The continuous emergence of magnetic flux brought in successively emerging of coronal loops and the slowly rising of an overlying horizontal filament threads. Sequentially, there appeared the deformation of the loops, the plasmoids ejection from the loop top, and pairs of loop brightenings and jet moving along the untwisting filament threads. All the observational results indicate there exist magnetic reconnection between the emerging loops and overlying horizontal filament threads, and it is the first example of two-sided-loop jets associated with ejected plasmoids and twisted overlying fields.
\end{abstract}


\section{Introduction}
Coronal jets, first discovered in soft X-ray on Yohkoh satellite (Shibata et al. 1992), are transient and collimated plasma flows. Coronal jets represent a group of impulsive events in different scales and develop in different layers of the solar atmosphere, including spicules, H$\alpha$ surges, Extreme ultraviolet (EUV) jets, and X-ray jets. Coronal jets always involve open field lines in coronal holes (CHs) or large-scale closed loops (Shibata et al. 1992; Schmieder et al. 1995; Rachmeler et al. 2010; Pariat et al. 2015), which may provide possible contributions to coronal heating and solar wind acceleration (Innes et al. 1997; Cirtain et al. 2007; Shibata et al. 2007; Pariat et al. 2009; Tian et al. 2014).

With the improvements of the temporal and spatial resolution of the telescopes, more and more new morphological features and physical characteristics of solar jets are revealed (see the recent reviews by e.g., Raouafi et al. 2016, and references therein), such as the composition of the cool and hot components (Zhang et al. 2000; Liu \& Kurokawa 2004; Jiang et al. 2007; Nishizuka et al. 2008), the recurrence from the same arch-base (Cirtain et al. 2007; Pariat et al. 2010; Innes et al. 2011; Jiang et al. 2013; Tian et al. 2014; Zhang et al. 2014b; Chen et al. 2015), and the untwisting motions around the spire (Patsourakos et al. 2008; Liu et al. 2011; Shen et al. 2011; Chen et al. 2012; Zhang et al. 2014a; Cheung et al. 2015; Lee et al. 2015; Liu et al. 2016). It is generally believed that coronal jets are always formed by magnetic reconnection between emerging flux and ambient pre-existing magnetic fields (Shibata et al. 1994; Raouafi et al. 2016).

According to the key mechanism of magnetic reconnection, coronal jets are classified into the standard model and the blowout model with two magnetic reconnection processes (Moore et al. 2010, 2013; Sterling et al. 2015, 2016). On the other hand, Shibata et al. (1994) proposed two types of coronal jets, namely the anemone type and two-sided-loop type, based on the different configurations of overlying magnetic fields. The former are single field-directed jets that occur between emerging flux and vertical/oblique magnetic fields, and the latter are bi-directional jets or loop brightenings that form between the emerging flux and the overlying horizontal magnetic fields (Yokoyama \& Shibata 1995, 1996). The scenario of two-sided-loop jets also predicts that the formation of some plasmoids (magnetic islands or blobs) are ejected in one direction accompanying with the bi-directional jets, and the plasmoids confine a cool dense chromospheric plasma. Furthermore, when there exist magnetic shear in the overlying fields, the vortex motion will appear around the plasmoids (Yokoyama \& Shibata 1996).

To our knowledge, only a few of the two-sided-loop jets have been reported, but the observations of plasmoids have never been detected in the cases (Alexander \& Fletcher 1999; Jiang et al. 2013). Can the rare two-sided-loop jets be detected frequently by observations with the high temporal and spatial resolution? In this paper, we report a example of successive two-sided-loop jets simultaneously observed in the chromosphere, transition region, and corona. The paper is organized as follows. Section 2 describes the observations; The main results are presented in Section 3; We give conclusions and discussion in Section 4.

\section{Observations and Data Analysis}
The two-sided-loop jets emanated from an emerging flux region (EFR) in NOAA active region (AR) 12681 on 2017 September 23. To analyse the jet evolution, we employ the EUV observations from Atmospheric Imaging Assembly (AIA: Lemen et al. 2012) onboard Solar Dynamics Observatory (SDO: Pesnell et al. 2012), the narrow-band slit-jaw images (SJIs) from Interface Region Imaging Spectrograph (IRIS: de Pontieu et al. 2014), and the H$\alpha$ filtergrams from New Vacuum Solar Telescope (NVST; Liu et al. 2014; Xiang et al. 2016) in the Fuxian Solar Observatory of China. We mainly use three AIA EUV channels, including 304~{\AA} (He II, $\sim$0.05 MK), 171~{\AA} (Fe IX, $\sim$0.6 MK), and  131~{\AA} (Fe VIII, $\sim$0.4 MK). Each AIA image covers the full disk (4096~$\times$ 4096 pixels) of the Sun and up to 0.5 $R_{sun}$ above the limb, with a pixel size of 0$\farcs$6 and a cadence of 12 seconds. The event is also scanned in the IRIS 1330 and 1400~{\AA} from 07:30 to 10:40 UT on September 23. Each SJI has the field of view (FOV) of $120^{''}$$\times$$120^{''}$, and its time cadence and spatial resolution are 20 seconds and 0$\farcs$332, respectively. The NVST data used in this study were obtained in H$\alpha$ 6562.8~{\AA} from 07:21 to 09:00 UT on September 23. The H$\alpha$ filtergrams have a spatial resolution of 0$\farcs$3, a cadence of 27 s, and a FOV of $151^{''}$$\times$$151^{''}$, covering AR 12681.

To check the magnetic field evolution of the EFR, we utilise magnetograms from the Helioseismic and Magnetic Imager (HMI: Scherrer et al. 2012), with a cadence of 45 seconds and pixel scale of 0$\farcs$6. In addition, the jets are also detected in the soft X-ray images of the X-ray Telescope (XRT; Golub et al. 2007; Kano et al. 2008) on the HINODE (Kosugi et al. 2007), and we mainly employ the Be\_thin images with a spatial resolution of $\sim$1$\farcs$0 pixel$^{-1}$ and a cadence of 90 seconds, though only part of the jets are captured due to the saturation of the XRT images. Finally, the images in different instruments are aligned by carefully comparing an obvious same feature (emerging loops, brightenings, and plasmoids), all the images in each wavelength are de-rotated with a reference image at $\sim$08:00 UT on September 23 by the routine DROT\_MAP of SolarSoft/IDL.

\section{Results}
\subsection{Magnetic flux emergence}
The overview of AR 12681 prior to the jets is shown in Figure 1. The AR consisted of a leading positive domain sunspot and following negative polarities (green and blue arrows in panel a). The EFR we selected involved the emerging loops (dotted boxes), and situated in the diffuse following polarities close to the polarity inversion line (PIL; the red dotted line in panel a) of the AR. Bundles of coronal loops in different heights and temperatures (arrows in panels b-c) straddled the PIL and covered the EFR. A filament (F; the yellow arrow in panel e) suspended along the PIL, and its north and south ends rooted in negative and positive polarities, respectively (blue and green arrows in panel e). In its close-up view, the middle part of F appeared as some faint slender threads (F1; the red arrow in panel f), connecting with dark portions on both sides (yellow arrows in panel f). In this paper, we mainly focus on F and F1, though another two filaments (F2 and F3) flew across F (white and black arrows in panels d-f).

Next, we focus on the evolution of the magnetic field in the EFR from one hour before the jet onset (Figure 2). In the EFR, the dominant negative polarity (N1) emerged obviously, and merged with the pre-existing other negative polarities in the south (blue arrows in panels a-d). A minor positive polarity (P1; green arrows in panels b-d) was diffuse and faint, and enclosed one isolated negative polarity (N2; the black arrow in panel a). N2 was fading after its emergence (panel d). The clear change of sizes for the P1, N1, and N2 in the magnetograms (panels a-d) likely indicates the emergence and cancellation of magnetic flux. Note that there was another negative polarity (N3; the white arrow in panel a) north to the EFR. Checked by the wavelet analysis, there are some periods of 300-400 s in the curves of magnetic fluxes for the P1, N1, and N2 (panels e-f). It is likely consistent with the 5-minute fluctuations of magnetic field due to the 5-minute p-mode oscillation in a sunspot (Lites et al. 1998).  Due to the fluctuations of the background magnetic fields, the weaker emergence and cancellation in the P1a-N2 region (blue dotted boxes in panels a-d) is not evident (panel e), but the strong emergence of N1 in the P1b-N1 region (white dotted boxes excluding blue dotted boxes in panels a-d) is very obvious (panel f). The N1 (blue) keeps growing, and has a rapid increasing phase from $\sim$08:10 to $\sim$08:30 UT. The magnetic flux of N1 ($\sim$ $10^{19}$ Mx) is stronger than that of N2 ($\sim$$10^{18}$ Mx) by one order of magnitude, which indicates the predominance of the emergence of N1 in the EFR. The jets just occurred during the rapid increasing phase of N1, and the times of two major jets are indicated by vertical dotted lines in panel f. The close temporal relationship shows that the continuous emergence of magnetic flux (N1 and P1) is intimately associated with the following jets.


\subsection{Emerging loops}
As the response for the magnetic flux emergence, small coronal loops emerged in the solar atmosphere (Figure 3). In H$\alpha$, the part of F over the EFR was still nearly invisible at $\sim$08:03 UT, but some filament threads (F1) became apparent and slowly rose (red arrows in panels a-c). On the other hand, small short loops (L1; white arrows in panels d-i) appeared in IRIS SJIs and AIA EUV images. Superposed with the contour of magnetic fields, the south footpoints of L1 clearly rooted at N1. The north footpoints of short L1 anchored in P1 (yellow dotted boxes in upper panels and Fig.2c) between N1 and N2 (blue and black arrows in panel b). Some minutes later, a new longer L1 appeared (panels f and i), and its south footpoints situated at the P1 (red dotted boxes in bottom panels and Fig.2c) north to N2.

Following its rise, the longer L1 became cusp in IRIS SJIs, AIA EUV and Hinode XRT images (yellow arrows in Fig.4b-d), and there appeared some brightenings in the EFR in H$\alpha$ (the white arrow in Fig.4a). Sequentially, the cusp of L1 disrupted, and a bright loop-like jet ejected from the top of L1 in IRIS 1400~{\AA}, AIA 171~{\AA}, and X-ray (green arrows in Figure 4f-h). The deformation of L1 and the occurrence of the loop-like jet are likely as a result of the approach and interaction between L1 and F1, and the jet nearly moved along the F1 (the red arrow in Fig.4e).



\subsection{Coronal jets}
In a minute, there emerged another new L1 after the first loop-like jet (Fig.5b-d). Then, the newly-emerged L1 again disrupted and spread into a bifurcated eruption that horizontally travelled in opposite directions as the two-sided-loop jets (see the e online animated version of this figure). The bi-directional jets were simultaneously seen in the chromosphere, transient region, and corona (white and yellow arrows in panels e-h). The northward jet moved along a small loop (L2; white arrows in panels f-g), shorter than $10^{''}$, in the form of loop brightenings. The L2 connected P1 and N3 (green and black arrows in panels f-g). The southward jet was much more attractive, and emanated from N1 (blue arrows in panels f-g). The southward jet propagated along F1 (red arrows in left panels) to remote positive polarities (the green arrow in Fig.1e) for a longer distance more than $30^{''}$. In X-ray, only the southward jet was distinguished (yellow arrows in panels h and l), and the L2 was occulted by the flaring core and overlying loops. Furthermore, the jets reoccurred intermittently, due to the successive emergence of L1. Another strong two-sided-loop jets happened at $\sim$08:24 UT as a pair of loop brightenings in the north and a southward jet (white and yellow arrows in panels j-k). Because the loop brightenings were short and weak, we mainly study on the southward strong jets.

Accompanying with the jets, some plasmoids/blobs escaped from the apex of the emerging L1, and five distinguished plasmoids are shown in Figure 6 (blue arrows). The third plasmoids (panels c-f) were tracked for a longer period, and other plasmoids were transient (panels a-b and h-i). The trajectories of the third plasmoids were superimposed on the H$\alpha$ filtergram (pluses in panel g), which clearly shows the third plasmoids first moved eastward, and then flowed southward, and finally disappeared near the F1 (the red arrow in panel g). It is similar to the vortex motion in the simulation of Yokoyama \& Shibata (1996). Along a horizonal dash line (S1; panel a), the evolution of the third plasmoids is clearly shown in the time-slice plots (panels j-l) in IRIS 1330~{\AA}, and AIA 171 and 304~{\AA}. The eastward initial speed of third plasmoids was $\sim$90 km s$^{-1}$ (blue dotted lines in panels j-l). In addition, it is also clear that the fast rising speed of L1 was $\sim$15.4 km s$^{-1}$ from $\sim$08:10 to $\sim$08:13 (the green dotted line in panel l), which is consistent with the emerging of the longer L1 in Fig.3i. The L1 emergence lasted from 08:00 to 08:30 UT, and the two intense patches (white arrows in panels j-l) are closely related with two major jets at $\sim$08:14 and $\sim$08:24 UT (green vertical lines in panels j-l).

The evolution of the F1 during the jet propagation is shown in the original and running-difference H$\alpha$ filtergrams (Figure 7). Before the jet onset, the twist was evident (the red arrow in panel a). The disconnection for the southward jets (yellow arrows in panels f-h) possibly indicates that the southward jets moved with the untwisting F1 (panels f-i). As the jets successively occurred, more and more filament threads of F1/F experienced an untwisting motion and became loose and straight at $\sim$08:30 UT (red arrows in panels b-e).

The evolution of the southward jets and F1 is displayed in time-slice plots in Figure 8. Along the jet path (S2 in Fig.7i), two major jets occurred at $\sim$08:14 and $\sim$08:24 UT (dashed vertical lines), respectively, consistent with the two strong patches in Fig.6j-l. The first jet was much more prominent, and propagated for more than 35 Mm with a speed of $\sim$120 km s$^{-1}$ (green dotted lines in panels a-e). The second jet had a faster velocity of $\sim$250 km s$^{-1}$ (blue dotted lines in panels a-d). Across the southern F1 (S3 in Fig.7i), a rise speed of $\sim$6 km s$^{-1}$ (the yellow dotted line in panel f) indicates that the southern F1 began to slowly upwards lift near the occurrence time of jets (the dashed vertical line in panel f). The transverse speed of $\sim$13 km s$^{-1}$ (green and blue dotted lines in panel f) confirms the untwisting motion of the southern F1. In the direction perpendicular to the F1 (S4 in Fig.7i) over the EFR, the rise speed of $\sim$6 km s$^{-1}$ (the green dotted line in panel g) indicates that F1 over the EFR slowly rose following the emergence of short L1 at the early stage, before the jet onset (the dashed vertical line in panel g).




\section{Conclusions and Discussion}
Combining with the high-quality observations from NVST, IRIS, and SDO, we report two-sided-loop jets on 2017 September 23. Beneath an active region filament, the continuous magnetic flux emergence in a small EFR induced the successively emerging loops (L1). At the early stage, a short L1 appeared, and some overlying filament threads (F1) began to slowly rise at a speed of $\sim$6 km s$^{-1}$. Then, a longer L1 appeared and had a rising velocity of $\sim$15 km s$^{-1}$. After the deformation (cusp) of the longer L1, the jets happened soon, and the southern F1 showed an upward expansion of $\sim$6 km s$^{-1}$ and a transverse motion of $\sim$13 km s$^{-1}$. For the jets, the northward jets moved in newly-formed short loops (L2) as loop brightenings, and the vigorous southward jets propagated along the south newly-formed filament loops. We analyse two major two-sided-loop jets, the speeds of two southward jets were $\sim$120, and $\sim$250 km s$^{-1}$, respectively. The faster speed of the latter jet is possibly caused by the decrease of magnetic field and rarefaction of plasma density after the propagation of the former one. The jets were also accompanied with plasmoids from the top of L1 and the untwisting motion of F1. These observations likely provide clear evidences of magnetic reconnection between emerging loops (L1) and twisted filament threads (F1).

We suggest a scenario of the two-sided-loop jets in Figure 9. The loops (L1; pink) emanate from the EFR, and their south and north ends root in the negative (N1) and positive (P1) polarities, respectively. L1 slowly rises and approaches to the twisted filament threads (F1; blue), which induces magnetic reconnection between L1 and F1 (the yellow cross). As a result, the north end of L1 connects to the north end of F1, in the form of short loop brightenings (L2; orange); another newly-formed filament loops (L3; red) links the south end of L1 with the south remote end of F1. The southward powerful jets (the green arrow) moved along newly-formed L3. As another production of magnetic reconnection, some plasmoids also escape eastward from the junction of L1 and F1 (the yellow cross), and then moved southward guided by F1 (yellow arrows in panel b). The continuous emergence of magnetic flux drives the successively emerging of L1, and results in the intermittent magnetic reconnection between L1 and F1. The intermittent magnetic reconnections between L1 and F1 form some more L2 and L3 (panel c).

The anemone jets and the two-sided-loop jets have a common nature of magnetic reconnection between the emerging flux and overlying fields. When the overlying filed is predominant horizontal, the resulting jets move in two different newly-formed closed loop structures (Yokoyama \& Shibata 1995, 1996). Yokoyama \& Shibata (1995) suggested that the two-sided-loop jets more likely appear as transient loop brightenings, not a necessary jet form. In this paper, because the junction point is close to the north ends of F1, the northward ejection moved along a newly-formed short loop (L2) and appeared as loop brightenings, while the intense southward ejection propagated along the longer untwisting filament threads in the form of jets. Hence, the two-sided-loop jets we studied here consist of pairs of loop brightenings and jets.

Up to date, there are only a few observational example of two-sided-loop jets (Alexander \& Fletcher 1999; Jiang et al. 2013; Tian et al. 2017). The two-sided-loop jets in Tian et al. 2017 were suggested to be caused by magnetic reconnection between adjacent filamentary threads, which is not accordant with the model of Yokoyama \& Shibata (1995,1996). Alexander \& Fletcher (1999) reported the two-sided-loop jets with the observations of Trace and Yohkoh, and Jiang et al. (2013) showed recurrent two-sided-loop jets that resulted from the magnetic reconnection between a emerging bipole and overlying transequatorial loops with the data. The simulation of two-sided-loop jets predicated the ejection of plasmoids (magnetic islands), and the plasmoids showed the vortex motion around the plasmoids when the overlying field had magnetic shear (Yokoyama \& Shibata 1996). To our knowledge, the plasmoids or overlying sheared fields in the two-sided-loop jet model has never been observed. Here, the two-sided-loop jets were observed simultaneously in the chromosphere (NVST), transition region (IRIS), and corona (AIA and XRT), and were associated with sheared fields (twisted filament threads) and plasmoids that have a vortex-like motion (Fig.6g). It is in agreement with the case of overlying sheared fields for the two-sided-loop jet model (Yokoyama \& Shibata 1995, 1996).

Coronal jets always display untwisting motions of helical structures. The untwisting motion always comes from the relaxation of the twist stored in the closed field, such as emerging bipole, small-scale flux ropes (Canfield et a. 1996; Jibben \& Canfield 2004; Pariat et al. 2009, 2010). In this case, the signals in AIA EUV images and IRIS SJIs are too weak to determine if there is the twist in the emerging L1. However, F1 had the twist stored in the filament threads. After the magnetic reconnection, the southward jets moved along the newly-formed filament loop and followed the untwisting motion of the filament threads. Therefore, the untwisting motion of jets in this event likely originates from the pre-exist overlying filament threads (F1). In addition, the untwisting velocity is nearly constant ($\sim$13 km s$^{-1}$) through the successive jets with different speeds (Fig.8f), which also confirms that the untwisting motion is the intrinsic property of the twisted filament threads. On the other hand, because there are bundles of coronal loops in different heights over the F1 and L1 (Fig.1b-c), the jet mass just flowed horizontally along newly-formed filament loops. Therefore, the successive jets are regarded as failed eruption, due to the strong confinement of the overlying large-scale loops.

The magnetic field strength of the L1 south footpoints is $B$ $\approx$ 68 $G$, and the initial density in the ambient corona is assumed to be $\rho \approx5 \times 10^{-19}$ $g$ $cm^{-3}$. Hence, the Alfv{\'e}n velocity is calculated by $v_A = B/{\sqrt{4 \pi \rho}}$ $\approx$ 260 km s$^{-1}$. According to the method in Yokoyama \& Shibata (1996), the speeds of plasmoids and jets are $v_{plasmoid} = 0.27 v_A$ $\approx$ 70 km s$^{-1}$ and $v_{jet} = 0.7 v_A$ $\approx$ 182  km  s$^{-1}$, which are comparable to that of the observations (90 and 120-250 km s$^{-1}$). Furthermore, the length of L1 was $\sim$15 Mm ($\sim$20$^{''}$), and that in the simulation of Yokoyama \& Shibata (1996) was $\sim$9.3 Mm. It seems that the two-sided-loop jets are always associated with shorted loops in the earlier emergence phase. We suggest that the two-sided-loop jets may occur as frequent as anemone jets, and they probably just show unresolved brightenings in low-resolution observations. It is a possible reason for the rarity of two-sided-loop jets. Further better observations will be necessary to detect more and more examples of two-sided-loop jets and probe their physical nature.


\acknowledgments
IRIS is a NASA Small Explorer mission developed and operated by LMSAL with mission operations executed at NASA ARC and major contributions to downlink communications funded by the NSC (Norway). SDO is a mission of NASA's Living With a Star Program, and SDO data and images are courtesy of NASA/SDO and the AIA and HMI science teams. The authors thank the NVST team for providing the data. This work is supported by grants NSFC 41274175, 41331068, 11303101, and 11603013, Shandong Province Natural Science Foundation ZR2016AQ16, and Young Scholars Program of Shandong University, Weihai, 2016WHWLJH07. H. Q. Song is supported by the Natural Science Foundation of Shandong Province JQ201710. This work is in memory of Dr. Yunchun Jiang who led R. Zheng into the gate of Solar Physics.

\clearpage

\begin{figure}
\epsscale{0.9} \plotone{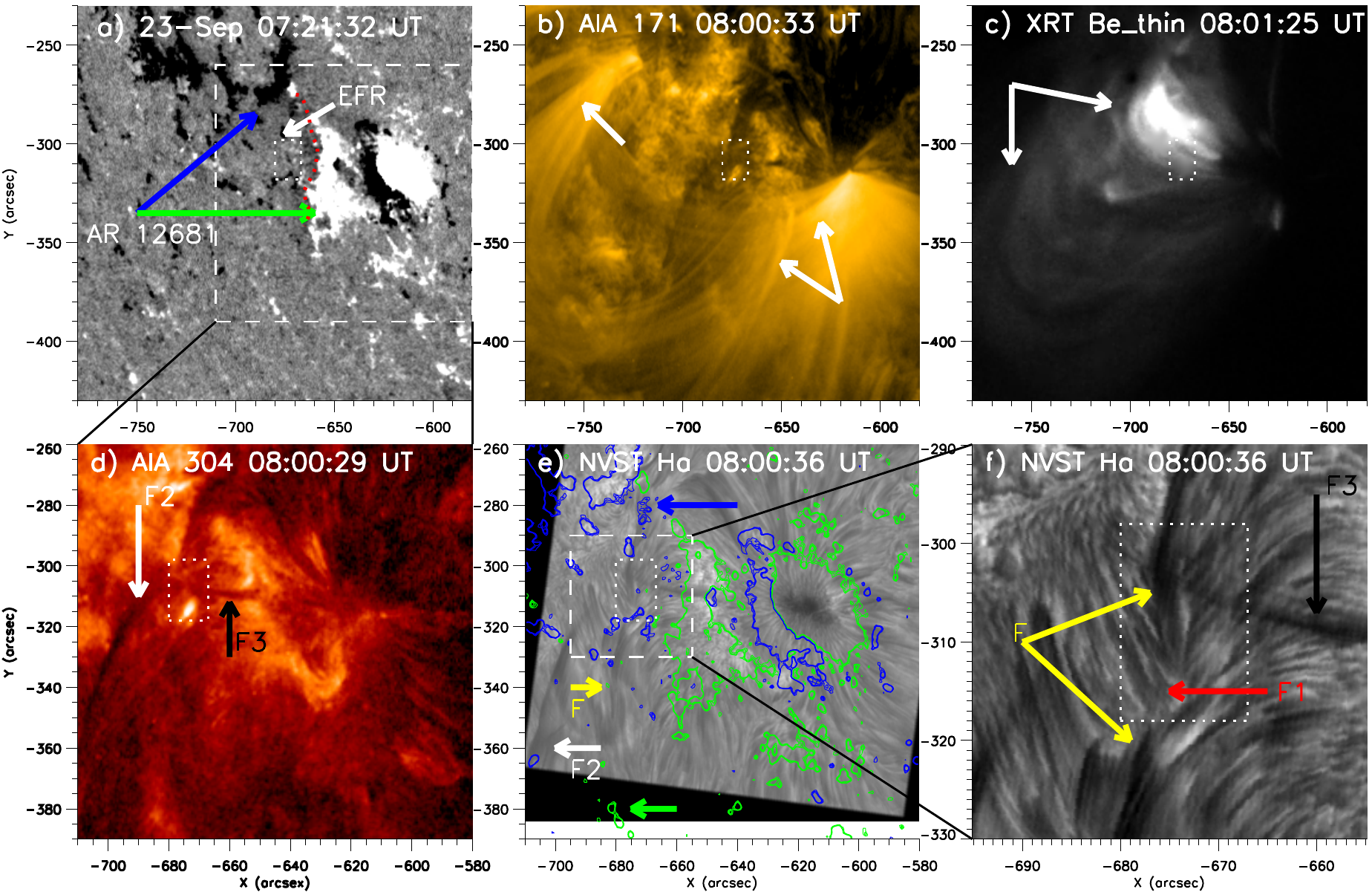}
\caption{Overview of AR 12681 and the EFR (dotted boxes) in an HMI line-of-sight magnetogram (a), in AIA 171 and 304~{\AA} (b and d), in XRT Be\_thin image (c), and in NVST H$\alpha$ filtergrams (e-f). The dashed box in (a) represents the FOV of (d-e), and the dashed box in panel (e) indicates the FOV of panel (f). The yellow arrows point out the filament (F) containing the interested filament threads (F1; the red arrow), and white arrows in panels (b-c) indicate the overlying multi-layer coronal loops. Contours of HMI longitudinal magnetic fields at 08:00:32 UT are superposed on panel (e) with positive (negative) fields in green (blue). The levels are 30, 40, and 50 G, respectively. The blue and green arrows in panel (e) show the magnetic polarities of the F ends. The white and black arrows in bottom panels show two overlying filaments (F2 and F3). The red dotted line in panel (a) represents the PIL.
\label{f1}}
\end{figure}

\clearpage

\begin{figure}
\epsscale{0.9}
\plotone{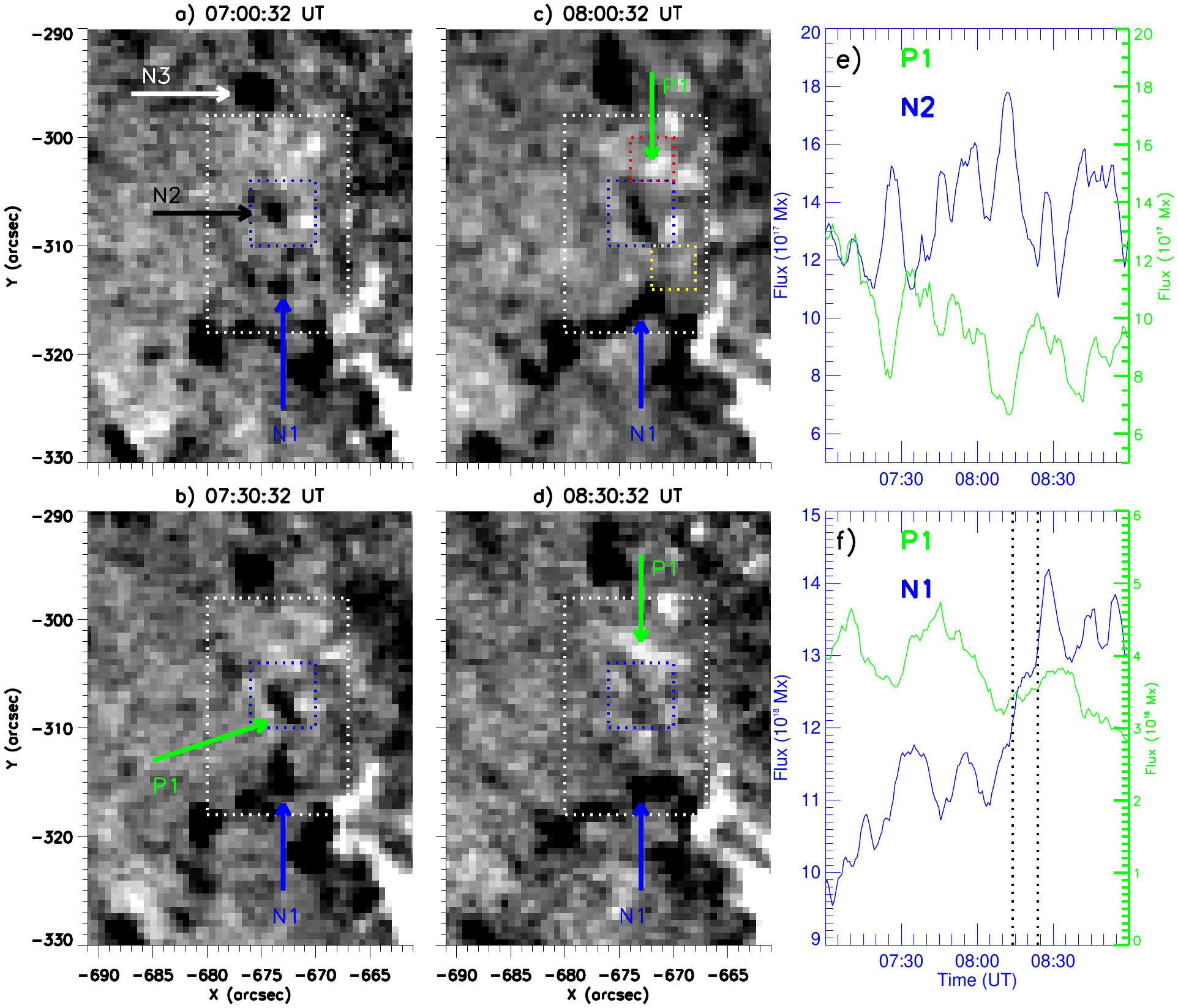}
\caption{(a-d) Evolution of the EFR (white dotted boxes) in HMI line-of-sight magnetograms with a range of (-30,30). The blue and green arrows indicate the emerging negative (N1) and positive (P1) polarities, respectively. The black and white arrows show another two negative polarities (N2 and N3). The yellow and red boxes indicate the north footpoints of emerging loops in Fig.3. (e) The plots of the positive (green) and negative (blue) magnetic flux for the region of P1-N2 (blue dotted boxes). (f) The curves of the positive (green) and negative (blue) magnetic flux for the EFR excluding P1-N2 region. The vertical lines mark the times of two major jets.
\label{f2}}
\end{figure}

\clearpage

\begin{figure}
\epsscale{0.9} \plotone{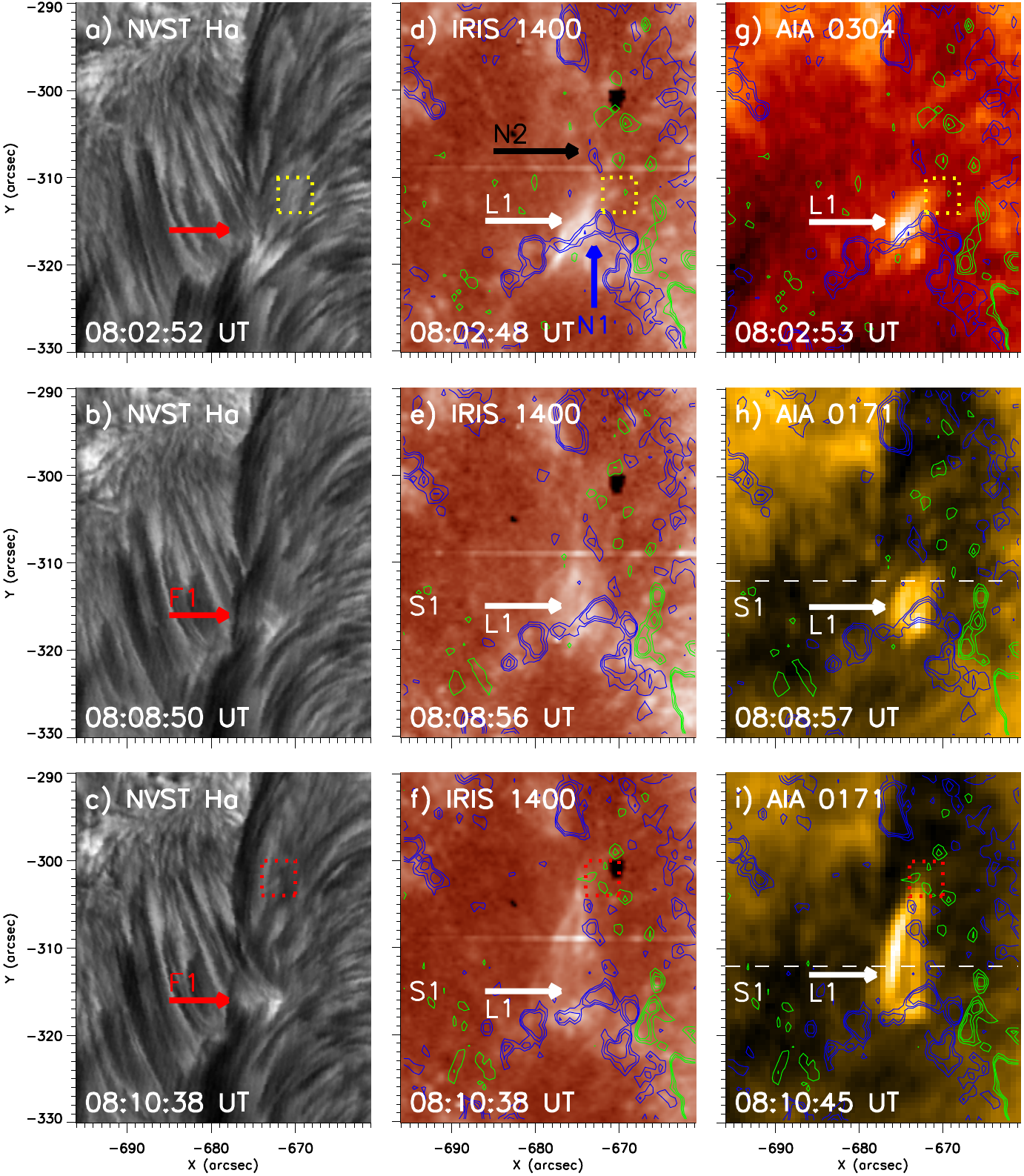}
\caption{The appearance of the filament threads (F1; red arrows) in NVST H$\alpha$ filtergrams and the emergence of small loops (L1; white arrows) in IRIS 1400 SJIs, and in AIA 304 and 171~{\AA}. Contours of HMI line-of-sight magnetic fields are superposed with positive (negative) fields in green (blue), and the levels are 15, 30, and 45 G, respectively. The yellow and red boxes indicate the north footpoints of L1. The dashed line (S1) in panels (h-i) is used to constructed the time-slice plots in Fig.6.
\label{f3}}
\end{figure}

\clearpage

\begin{figure}
\epsscale{0.9} \plotone{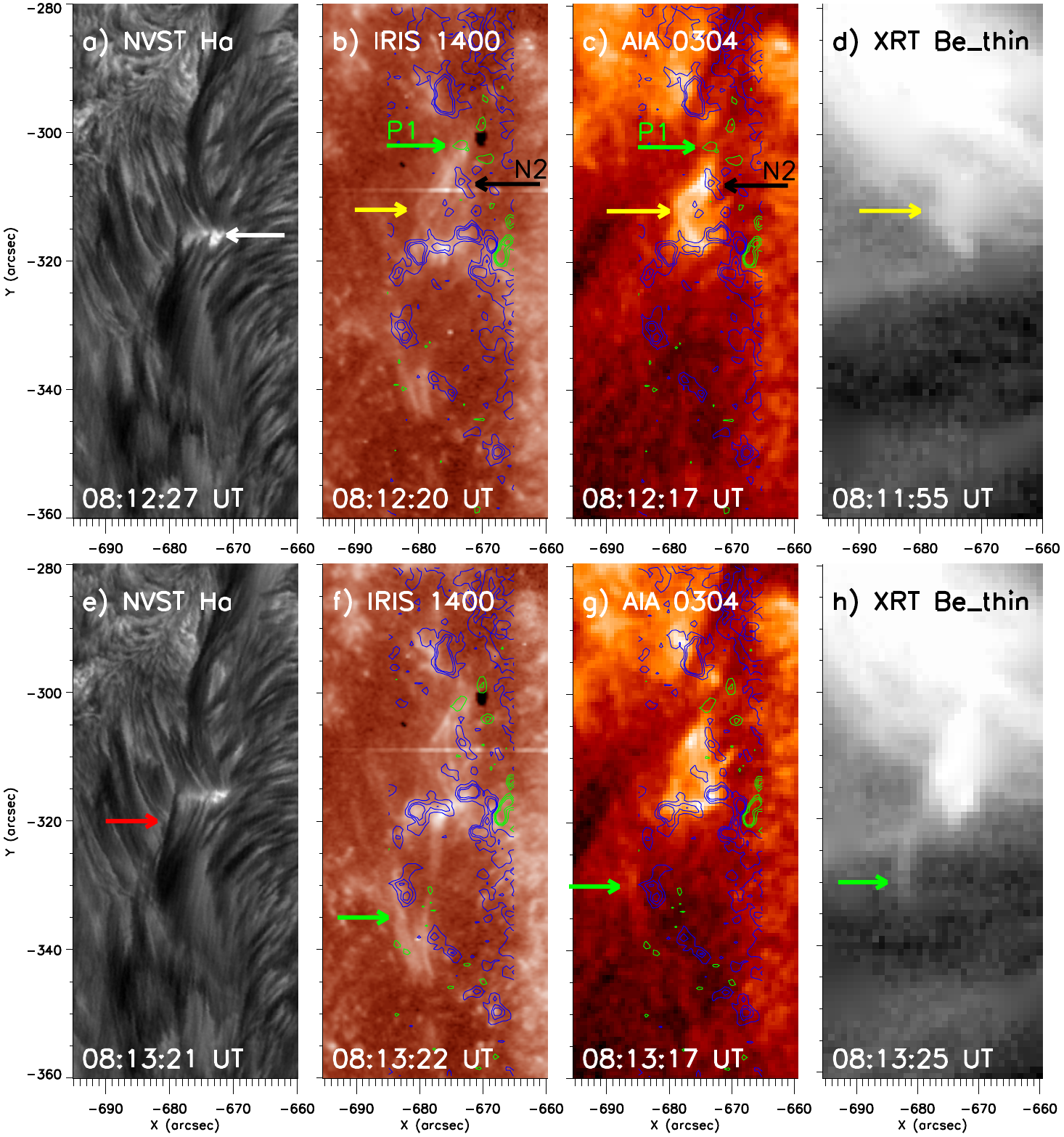}
\caption{The cusp of L1 (yellow arrows) and the loop-like jet (green arrows) and the overlying filament threads (the red arrow) in NVST H$\alpha$ filtergrams, IRIS 1400 SJIs, AIA 304 and 171~{\AA}, and XRT Be\_thin images. The white arrow in panel (a) indicates the brightenings before the jet onset. Contours of HMI line-of-sight magnetic fields are superposed with positive (negative) fields in green (blue), and the levels are 15, 30, and 45 G, respectively.
\label{f4}}
\end{figure}

\clearpage

\begin{figure}
\epsscale{0.72} \plotone{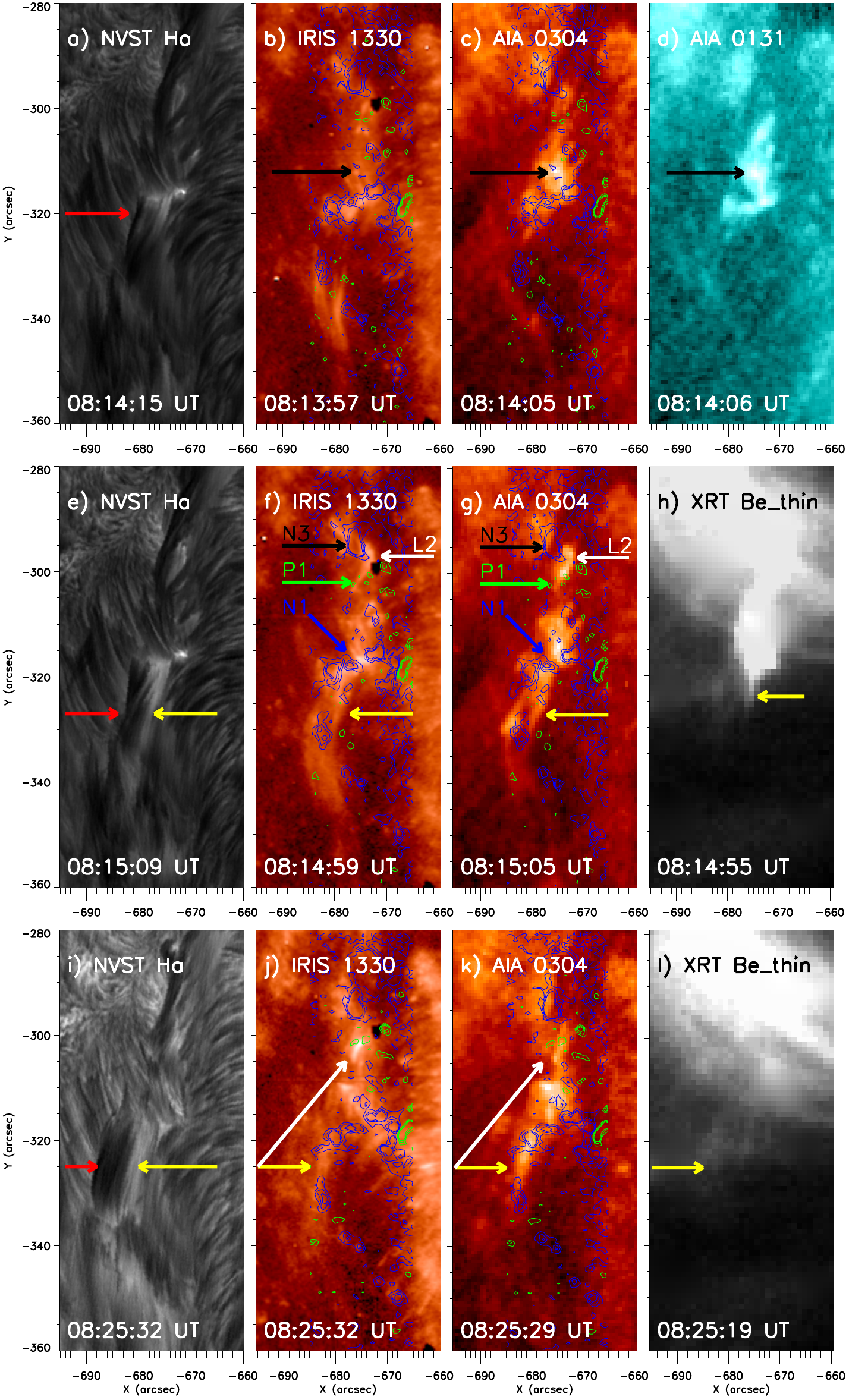}
\caption{Following the emergence of a new L1 (black arrows), the loop brightenings (white arrows), the untwisting F1 (red arrows), and southward jets (yellow arrows) in NVST H$\alpha$ filtergrams, IRIS 1330 SJIs, AIA 304 and 131~{\AA}, and XRT Be\_thin images. Contours of HMI line-of-sight magnetic fields are superposed with positive (negative) fields in green (blue), and the levels are 15, 30, and 45 G, respectively.
\label{f5}}
\end{figure}

\clearpage

\begin{figure}
\epsscale{0.9} \plotone{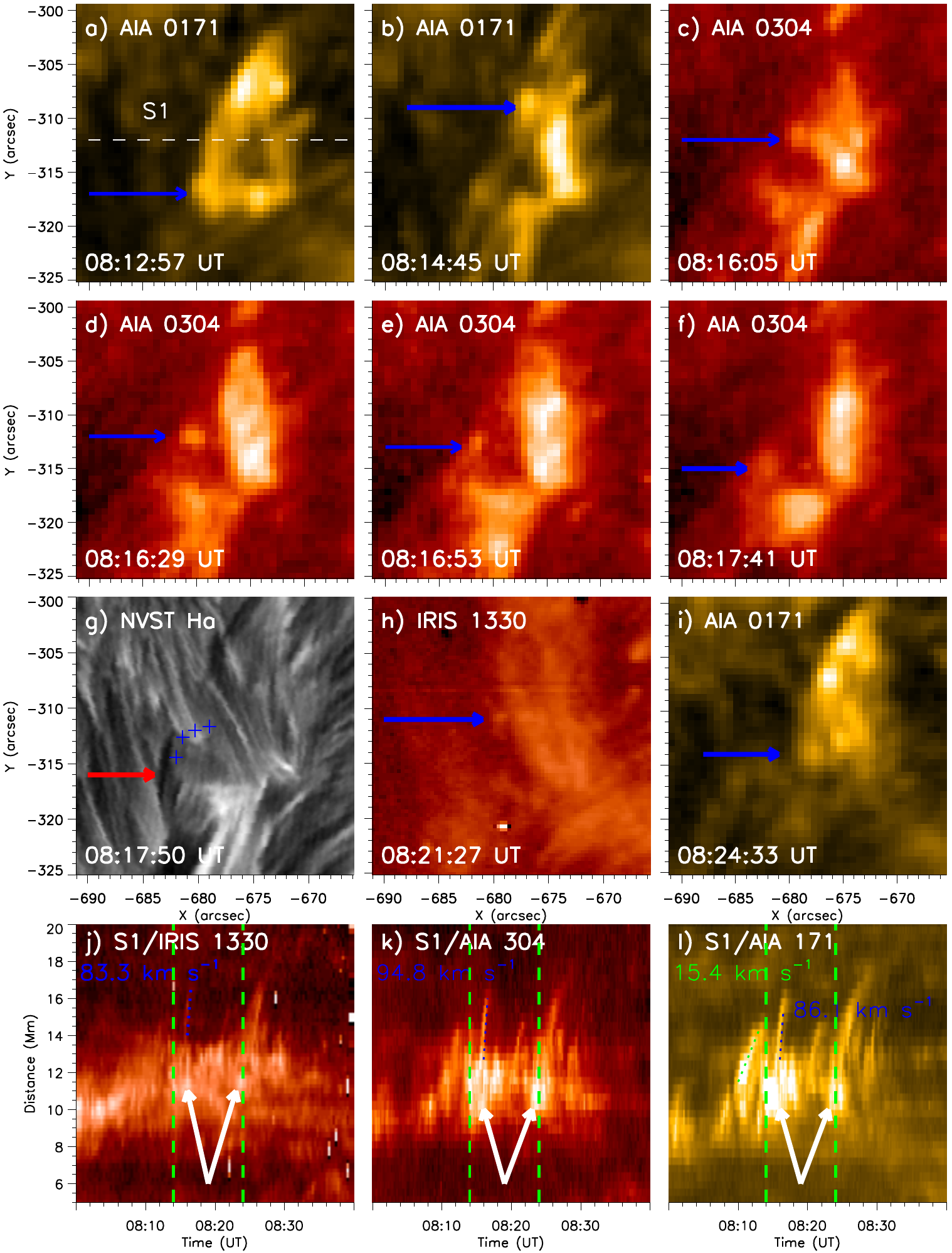}
\caption{(a-i) The successive plasmoids (blue arrows) associated with the jets. The trajectories (blue pluses) of the plasmoids in panels (c-f) show a vortex-like motion close to the filament threads (the red arrow in panel g). (j-l) The time-slice plots along the S1 (dashed lines in panels (e-f)). The green dotted line shows the initial rise of the L1, and the blue dotted lines indicate the release of the plasmoids. The white arrows point out two strong emergences.
\label{f6}}
\end{figure}

\clearpage

\begin{figure}
\epsscale{0.9} \plotone{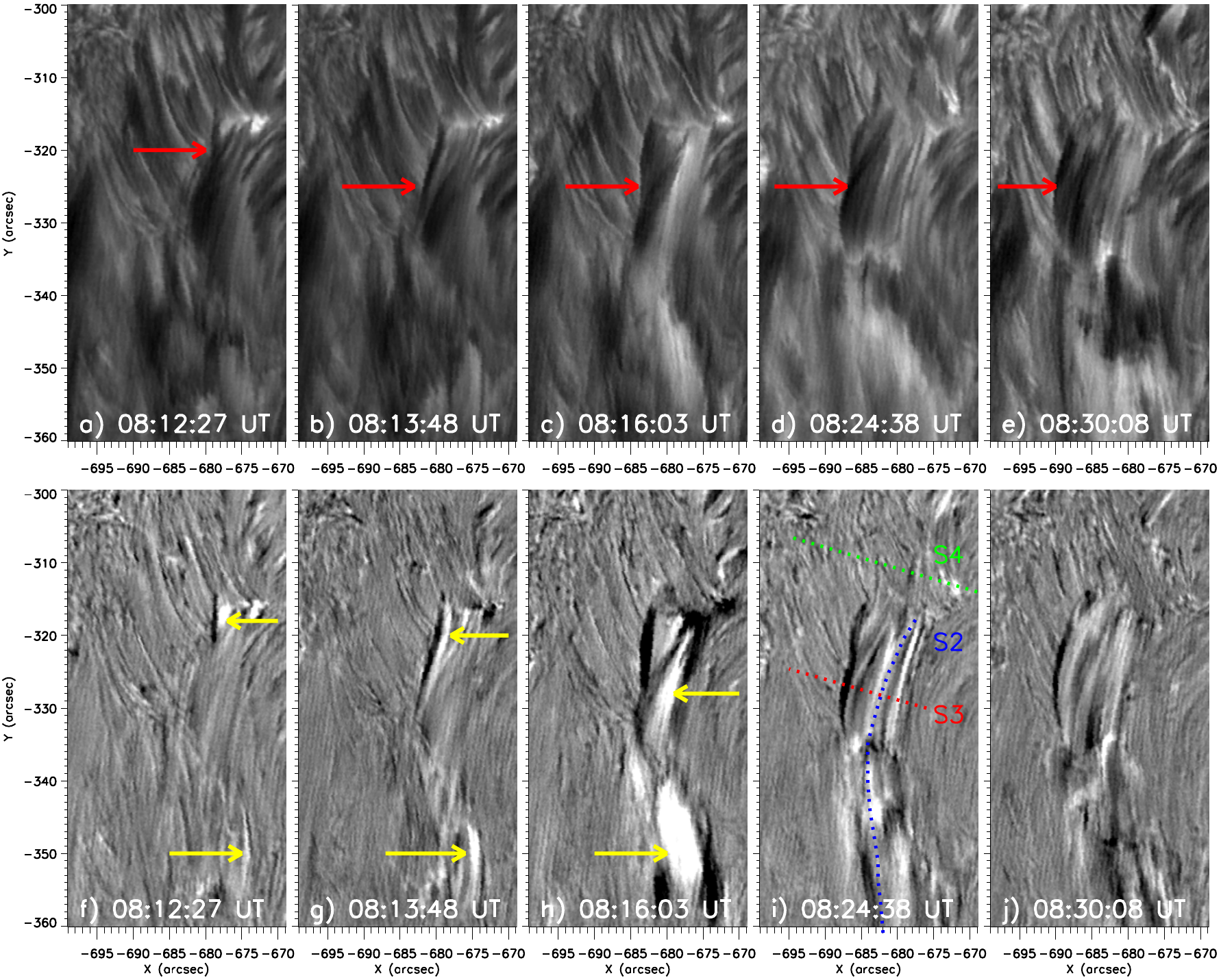}
\caption{The evolution of the F1 (red arrows) during the jet propagation (yellow arrows) in the original (a-e) and running-difference (f-j) H$\alpha$ filtergrams. The dotted lines (S2-S4) in panel (i) are used to constructed the time-slice plots in Fig.8.
\label{f7}}
\end{figure}

\clearpage

\begin{figure}
\epsscale{0.9} \plotone{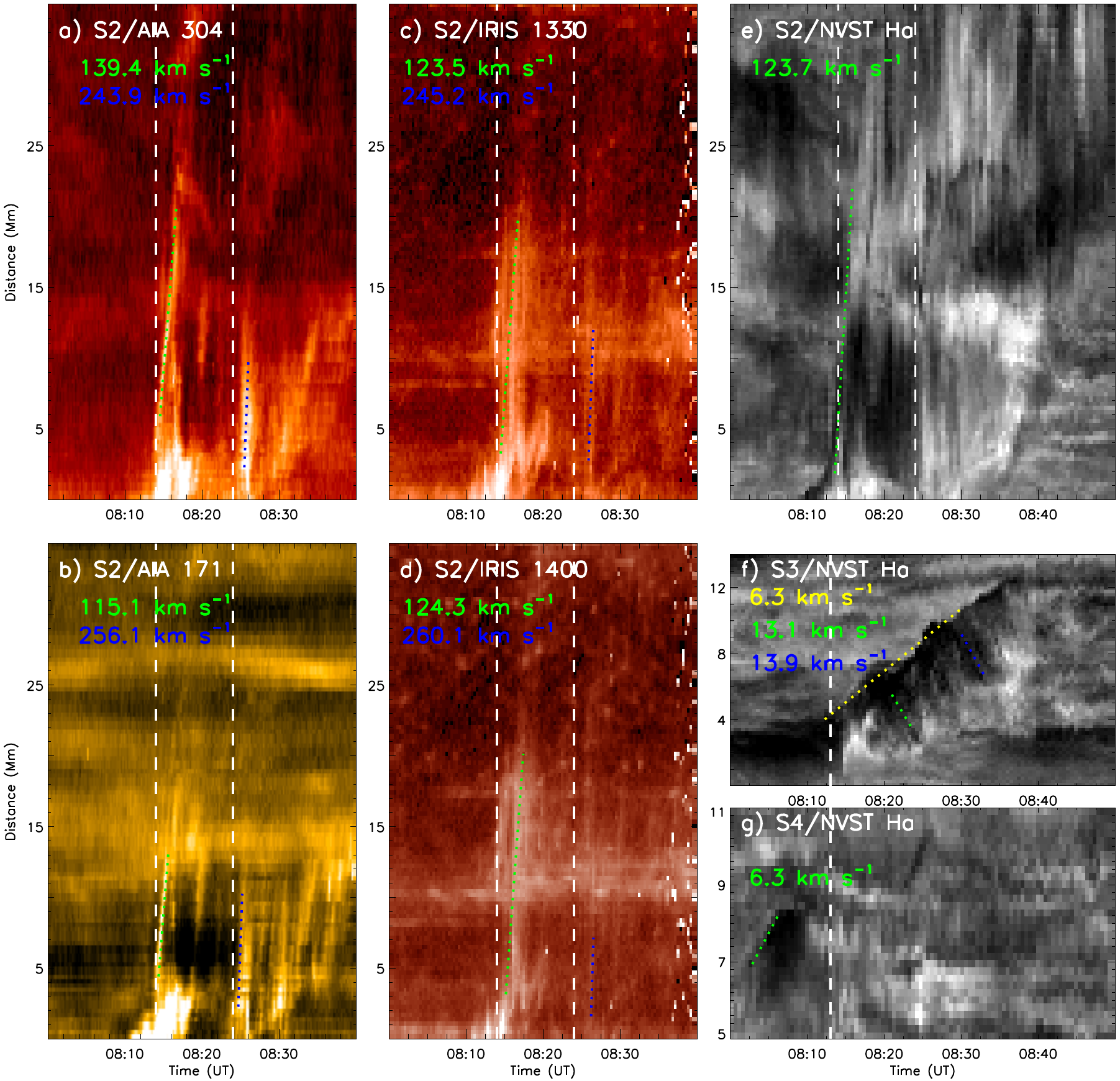}
\caption{The time-slice plots for the propagation of the southward jets along S2 (a-e), the untwisting motion of the F1 along S3 (f), and the rise of F1 before the jet onset along S4 (g). The vertical dashed lines in panels (a-e) show the times of two major jets, and the jet speeds are obtained by the green and blue dotted lines in panels (a-e). The vertical dashed line in panels (f-g) indicates the beginning time of the first loop-like jet. In panel (f), the untwisting speeds are calculated by the green and blue dotted lines, and the eastward expansion velocity of the untwisting F1 is measured by the yellow dotted line. In panel (g), the ascent speed of the F1 before the jet onset is determined by the green dotted line.
\label{f8}}
\end{figure}

\clearpage

\begin{figure}
\epsscale{0.9} \plotone{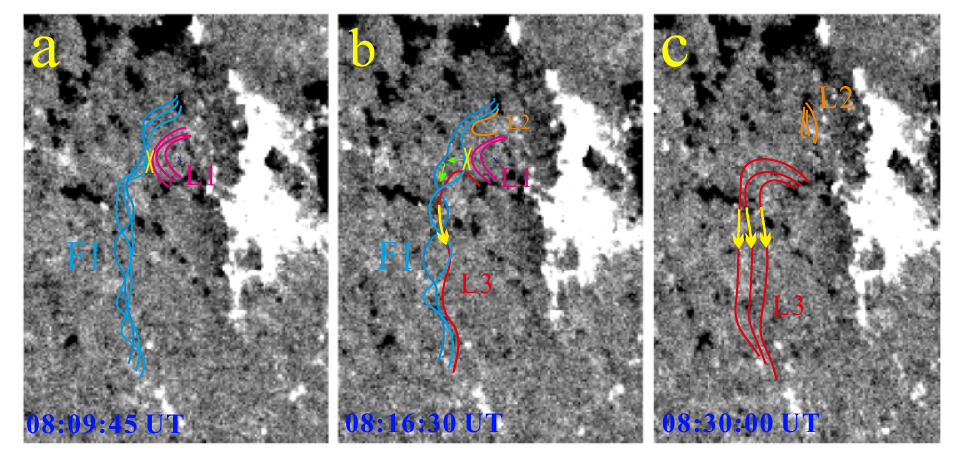}
\caption{The schematic representation showing the scenario of magnetic reconnection between the emerging loops (L1; pink) and the twisted filament threads (F1; blue). As a result, the plasmoids (green arrows) released from the junction (yellow cross) of F1 and L1, and there newly formed loop brightening (L2; orange) in the north and the longer filament loops (L3) guiding the southward jets (yellow arrows).
\label{f9}}
\end{figure}


\begin{thebibliography}{}

\bibitem[Alexander \& Fletcher (1999)]{alex99} Alexander, D., \& Fletcher, L. 1999, \solphys, 190, 167

\bibitem[Brueckner \& Bartoe (1983)]{brue83} Brueckner, G. E. \& Bartoe, J.-D. F. 1983, \apj, 272, 329

\bibitem[Canfield et al. (1996)]{canf96} Canfield, R. C., Reardon, K. P., Leka, K. D., et al. 1996, \apj, 464, 1016

\bibitem[Chen et al. (2012)]{chen12} Chen, H.-D., Zhang, J., \& Ma, S.-L. 2012, RAA, 12, 573
\bibitem[Chen et al. (2015)]{chen15} Chen, J., Su, J., Yin, Z., et al. 2015, \apj, 815, 71

\bibitem[Cheung et al. 2015]{cheung15} Cheung, M. C. M., De Pontieu, B., Tarbell, T. D., et al. 2015, \apj, 801, 83

\bibitem[Cirtain et al. (2007)]{cirt07} Cirtain, J. W., Goloub, L., Lundquist, L., et al. 2007, Science, 318, 1580

\bibitem[de Pontieu et al. (2014)]{depon14} de Pontieu, B., Title, A. M., Lemen, J. R., et al. 2014, \solphys, 289, 2733

\bibitem[Dere et al. (1989)]{dere89} Dere, K. P., Brueckner, G. E., \& Bartoe, J.-D. F. 1989, \solphys, 123, 41

\bibitem[Golub et al. (2007)]{golub07} Golub, L., DeLuca, E., Austin, G., et al. 2007, \solphys, 243, 63

\bibitem[Innes et al. (1997)]{innes97} Innes, D. E., Inhester, B., Axford, W. I., \& Wilhelm, K. 1997, Nature, 386, 329

\bibitem[Jiang et al. (2013)]{jiang13} Jiang, Y., Bi, Y., Yang, J., et al. 2013, \apj, 775, 132

\bibitem[Jiang et al. (2007)]{jiang07} Jiang, Y. C., Chen, H. D., Li, K. J., Shen, Y. D., \& Yang, L. H. 2007, \aap, 469, 331

\bibitem[Jibben \& Canfield (2004)]{jibben04} Jibben, P., \& Canfield, R. C. 2004, \apj, 610, 1129

\bibitem[Kano et al. (2007)]{kano} Kano, R., Sakao, T., Hara, H., Tsuneta, S., et al. 2007, \solphys, 249, 263

\bibitem[Kosugi et al. (2007)]{kosugi} Kosugi, T., Matsuzaki, K., Sakao, T., Shimizu, T., et al. 2007, \solphys, 243, 3

\bibitem[Lee et al. (2015)]{lee15} Lee, E. J., Archontis, V., \& Hood, A. W. 2015, \apjl, 798, L10

\bibitem[Lemen et al. (2012)]{lemen12} Lemen, James R., Title, Alan M., Akin, David J., Boerner, Paul F. et al. 2012, \solphys, 275, 17

\bibitem[Lites et al. (1998)]{lites98} Lites, Bruce W., Thomas, John H., Bogdan, Thomas J., \& Cally, Paul S. 1998, \apj, 497, 464

\bibitem[Liu et al. (2011)]{liu11} Liu, C., Deng, N., Liu, R., et al. 2011, \apjl, 735, L18

\bibitem[Liu et al. (2016)]{liu16} Liu, J., Wang, Y., Erd{\'e}lyi, R., Liu, R., et al. 2016, \apjl, 833, L150
\bibitem[Liu \& Kurokawa (2004)]{liu04} Liu, Y., \& Kurokawa, H. 2004, \apj, 610, 1136
\bibitem[Liu et al. (2014)]{liu14} Liu, Z., Xu, J., Gu, B.-Z., et al. 2014, RAA, 14, 705


\bibitem[Moore et al. (2010)]{moore10} Moore, R. L., Cirtain, J. W., Sterling, A. C., \& Falconer, D. A. 2010, \apj, 720, 757
\bibitem[Moore et al. (2013)]{moore13} Moore, R. L., Sterling, A. C., Falconer, D. A., \& Robe, D. 2013, \apj, 769, 134

\bibitem[Nishizuka et al. (2008)]{nishi08} Nishizuka, N., Shimizu, M., Nakamura, T., et al. 2008, \apjl, 683, L83


\bibitem[Pariat et al. (2009)]{pari09} Pariat, E., Antiochos, S. K., \& DeVore, C. R. 2009, \apj, 691, 61

\bibitem[Pariat et al. (2010)]{pari10} Pariat, E., Antiochos, S. K., \& DeVore, C. R. 2010, \apj, 714, 1762

\bibitem[Pariat et al. (2015)]{pari15} Pariat, E., Dalmasse, K., DeVore, C. R., Antiochos, S. K., \& Karpen, J. T. 2015, \aap, 573, 130

\bibitem[Patsourakos et al. (2008)]{pats08} Patsourakos, S., Pariat, E., Vourlidas, A., et al. 2008, \apjl, 680, 73

\bibitem[Pesnell et al. (2012)]{pesnell12} Pesnell, W. Dean, Thompson, B. J., Chamberlin, P. C. et al. 2012, \solphys, 275, 3

\bibitem[Rachmeler et al. (2010)]{rach10} Rachmeler, L. A., Pariat, E., DeForest, C. E., Antiochos, S. K., \& T{\"o}r{\"o}k, T. 2010, \apj, 715, 1556

\bibitem[Raouafi et al. (2016)]{raou16} Raouafi, N. E., Patsourakos, S., Pariat, E., et al. 2016, Space Science Review, 201, 1

\bibitem[Shen et al. (2012)]{shen12} Shen, Y., Liu, Y., Su, J., \& Deng, Y. 2012, \apj, 745, 164
\bibitem[Shen et al. (2011)]{shen11} Shen, Y., Liu, Y., Su, J., \& Ibrahim, A. 2011, \apjl, L735, 43

\bibitem[Scherrer et al. 2012]{sche12} Scherrer, P. H., Schou, J., Bush, R. I., Kosovichev, A. G., et al. 2012, \solphys, 275, 207

\bibitem[Schmieder et al. (1995)]{schm95} Schmieder, B., Shibata, K., van Driel-Gesztelyi, L., \& Freeland, S. 1995, \solphys, 156, 245

\bibitem[Shibata et al. 1992]{shib92} Shibata, K, Ishido, Y., Acton, L. W. et al. 1992, Publications of the Astronomical Society of Japan, 44, 173

\bibitem[Shibata et al. 2007]{shib07} Shibata, K, Nakamura, T, Matsumoto, T. et al. 2007, Science, 318, 1591

\bibitem[Shibata et al. (1994)]{shiba94} Shibata, K., Nitta, N., Matsumoto, R., et al. 1994, in X-ray Solar Physics from Yohkoh, ed. Y. Uchida et al. (Tokyo: Universal Academy Press), 29

\bibitem[Sterling et al. (2016)]{ster16} Sterling, A. C., Moore, R. L., Falconer, D. A., et al. 2016, \apj, 821, 100
\bibitem[Sterling et al. (2015)]{ster15} Sterling, A. C., Moore, R. L., Falconer, D. A., \& Adams, M. 2015, Nature, 523, 437

\bibitem[Tian et al. (2014)]{tian14} Tian, H., DeLuca, E. E., Cranmer, S. R., et al. 2014, Science, 346, 1255711

\bibitem[Tian et al. (2017)]{tian17} Tian, Z., Liu, Y., Shen, Y., et al. 2017, \apj, 845, 94

\bibitem[Xiang et al. (2016)]{xiang16} Xiang, Y., Liu, Z., \& Jin, Z. 2016, New Astronomy, 49, 8

\bibitem[Yokoyama \& Shibata (1995)]{yoko95} Yokoyama, T., \& Shibata, K. 1995, Nature, 375, 42
\bibitem[Yokoyama \& Shibata (1996)]{yoko96} Yokoyama, T., \& Shibata, K. 1996, PASJ, 48, 353

\bibitem[Zhang et al. (2000)]{zhang00} Zhang, J., Wang, J., \& Liu, Y. 2000, \aap, 361, 759

\bibitem[Zhang et al. (2014a)]{zhang14a} Zhang, Q., \& Ji, H. 2014a, \aap, 561, 134
\bibitem[Zhang et al. (2014b)]{zhang14b} Zhang, Q., \& Ji, H. 2014b, \aap, 567, 11

\end{thebibliography}
\end{document}